\documentclass{article}
\usepackage{emulateapj,pstricks,apjfonts}

\long\def\***#1{{\sc #1}}

\begin{document}

\lefthead{RXTE OBSERVATIONS OF X-RAY NOVA GS~1354--644}
\righthead{REVNIVTSEV ET AL.}

\submitted{ApJ accepted, scheduled for Feb. 2000, v.530}

\title{RXTE OSERVATIONS OF AN OUTBURST OF RECCURENT X-RAY NOVA GS~1354--644}

\author{Mikhail G. Revnivtsev$^{1,2}$, Konstantin N. Borozdin$^{2,1}$, 
William C. Priedhorsky$^2$, Alexey Vikhlinin$^3$}

\affil{$^{1}$ -- Space Research Institute, Moscow, Russia\\
$^{2}$ -- Los Alamos National Laboratory, Los Alamos, NM 87545, USA\\
$^3$ -- Harvard-Smithsonian Center for Astrophysics, Cambridge, MA 02138, USA}

\begin{abstract}
We present the results of Rossi X-ray Timing Explorer observations of
GS~1354--644 during a modest outburst in 1997-1998. 
The source is one of a handful of black hole X-ray transients 
that are confirmed to be recurrent in X-rays.
A 1987 outburst of the same source observed by Ginga was much 
brighter, and showed a high/soft spectral state.  
In contrast the 1997-1998 outburst showed a low/hard
spectral state.  Both states are typical for black hole binaries.
The RXTE All Sky Monitor observed an outburst duration of
150 to 200 days.  PCA and HEXTE observations covered $\sim$70 days 
near the maximum of the light curve and during the flux decline.  
Throughout the observations, the spectrum can be approximated by
Compton upscattering of soft photons by energetic electrons.
The hot electron cloud has a temperature kT $\sim$30 keV and optical 
depth $\tau \sim$4--5. To fit the data well an additional 
iron fluorescent line and reflection component are required,
which indicates the presence of optically 
thick cool material, most probably in the outer part 
of the accretion disk. Dramatic fast variability was observed, 
and has been analyzed in the context of a shot noise model.
The spectrum appeared to be  
softest at the peaks of the shot-noise variability.
The shape of the power spectrum was typical for black hole systems
in a low/hard state.  We note a qualitative difference in the shape 
of the dependence of fractional variability on energy, when we
compare systems with black holes and with neutron stars.
Since it is difficult 
to discriminate these systems on spectral grounds, at least 
in their low/hard states, this new difference might be important.
\end{abstract}

\section{Introduction}

A modest X-ray outburst from the recurrent transient X1354--644 was
detected by the All Sky Monitor (ASM) aboard the Rossi X-ray Timing 
Explorer (RXTE) around 1 November 1997 (\cite{asm}). 
The flux from the source as measured by the ASM in the 2-12 keV 
energy band, rose gradually to 40-50 mCrab in mid-November, 
stayed at this level for about a month, then started a slow decline. 
The maximum hard X-ray flux observed by BATSE (\cite{batse}) and HEXTE
(\cite{hexte_gs}) appeared to be around 150 mCrab, indicating a hard 
spectrum.  HEXTE detected an exponential spectral cutoff at higher 
energies.

Castro-Tirado (1997) obtained B- and R-band images following 
the detection of the source in X-rays. The optical counterpart BW Cir 
was found to be in a bright state. Radio (\cite{fender}) 
and infrared (\cite{soria}) emission was also detected during 
the X-ray outburst. Spectroscopic observations obtained in January 
1998, reveal a strong H-alpha line in emission, with a profile
that varied from double- to single-peaked over 
three nights, reminiscent
of the black hole microquasar GRO~1655-40 during its 1996 outburst, 
and strongly suggesting an accretion disk origin (\cite{bux}).

The X-ray transient source X1354--644 belongs to the class of X-ray 
binaries known as X-ray Novae (e.g. \cite{sun94}; \cite{tsh96}).
All sources in this class are assumed to be recurrent, but 
only a few have been observed in more than one X-ray outburst.
In case of X1354--644, thanks to observations of a common 
optical counterpart (\cite{peder87,castro}), one can reliably
identify the 1997 X-ray event with one observed by Ginga 
in 1987 as GS~1354--64 (\cite{mak87}, \cite{kit90}). 
It is quite probable that X-ray emission from the same source, 
under different names, was detected also 
in 1967 (\cite{har67}, \cite{fran71}), in 1971-1972 (\cite{mark77}), 
and in 1974-1976 (\cite{sew76}; \cite{wood84}).  
Those earlier observations however are more disputable 
because of relatively poor ($\sim$ degrees) localization 
for the X-ray source and the lack of 
counterpart detections at other wavelengths.

Different outbursts of the source varied substantially
in peak flux.
The most prominent outburst from this area of sky was detected in 1967 
as Cen~X-2 (\cite{har67}).  It was the first X-ray transient 
found, and still one of the brightest Galactic X-ray sources known. 
Its association with GS~1354--644 is however questionable 
because of poor position determination for Cen X-2, and because 
the 1967 outburst luminosity would have been much higher than 
Eddington.  The peak fluxes detected by Ginga
in 1987 (\cite{kit90}) and by OSO-7 in 1971-1972 (\cite{mark77})
were about 2 orders of magnitude weaker than Cen~X-2. 
The maximum flux detected by RXTE in 1997 
was about 3 times lower than in 1987 outburst.
 
The energy spectra were also quite different for the two outbursts
(1987 and 1997). The spectrum measured with Ginga in 1987 
was composed of a strong soft component below 10 keV and 
a power law at higher energies, 
which is typical for black-hole binaries in their ``high''
spectral state.  In contrast, the spectrum
detected by RXTE in the current outburst is typical of the
``low'' spectral state of Galactic X-ray binaries.

In this paper we present an analysis of the observations of
GS~1354--644 during its 1997-1998 outburst based on RXTE data.
An overview of observations and our data reduction approach are presented
in section 2. Our timing analysis is presented in \S3, and spectral
analysis in \S4.  We compare our results with other sources and
with some theoretical models in section 5.
We summarize our results in \S6.

\section{Observations and data reduction}

The Rossi X-ray Timing Explorer satellite (\cite{rxte})   
has two coaligned spectrometers the PCA and HEXTE, with narrow fields 
of view, as well as an All Sky Monitor (ASM).  
PCA and HEXTE together provide broad band spectral coverage 
in the energy range from 3 to $\sim 200$ keV.
The ASM tracks the long term behavior of sources in the 2-12 keV 
energy band. The data for our analysis have been obtained 
from the RXTE Guest Observer Facility at GSFC, which is a part of
High Energy Astrophysics Science Archive Research Center (HEASARC). 

The reduction of PCA and HEXTE data was performed with the standard 
{\it ftools} package. To estimate the PCA background we applied 
the $L7/240$ background model, which takes into account various
particle monitors and the SAA history, for the 17 November 1998 
observation, and the $VLE$ (Very Large Events) -based
model for other observations.

We used PCA response matrix v.3.3 (Jahoda 1998 a, b).
Analysis of the Crab nebula spectra confirmed
that the systematic uncertainties of the matrix were less 
than 1\% in the 3-20 keV energy band.  Uncertainties in 
the response, and a sharp decrease of PCA effective area below
3 keV, makes it hard to accurately measure low-energy spectral 
absorption. We have used PCA data only in the 3-20 keV range, 
and added 1\% systematic error for each PCA channel 
to account for residual uncertainties in the spectral 
response. All spectra were corrected for dead-time as per 
Zhang \& Jahoda (1996).

Response matrix v.2.6 was used to fit the HEXTE spectra
(\cite{hexte}). The background value for each 
cluster of HEXTE detectors was estimated from adjacent
off-source observations. An upper energy limit for the analysis
was defined according to the brightness of the source.  Typically,
we did not consider data at energies higher than $\sim 150-200$ keV,
where background subtraction uncertainties became unacceptably large. 
A dead time correction was applied to all observations.

The pointed RXTE observations are summarized in Table \ref{obslog}.  
The 1997 PCA and HEXTE observations were carried out near 
outburst maximum and adjacent rise and decline phases. 
During the observation of Nov 1998 the source was much dimmer,
probably at or approaching quiescence.

\section{Variability}

\subsection{Light curve of the outburst}

The 1997 light curve showed a triangular profile, possibly with 
a short plateau near maximum, similar to that observed 
from other X-ray transients (e.g. \cite{loch94}; \cite{har94}).
For general light curve morphologies see \cite{chen97}.

The light curve of the source measured by the ASM 
in the 2-12 keV energy band is presented in Fig.1. 
When approximated by an exponential function, 
the rise time parameter is $20 \pm 2$ days 
and decay time parameter around 40 days. 
Flux measurements by the PCA in the 2-30 keV energy band and by
HEXTE in the 20-100 keV energy band are shown at the same Figure. 
There is some evidence for a secondary maximum or ``kick'', 
typical for outbursts of X-ray Novae 
(e.g.  \cite{sun94}; review in \cite{tsh96}).
The light curve of the previous outburst of this source, 
tracked by the Ginga ASM, had a plateau and a decline with
time scales around 60 days (\cite{kit90}).  The peak flux of 
the 1997-1998 outburst in the 1-10 keV band extrapolated from PCA 
measurements was $\sim1.1 \times 10^{-9}$ erg s$^{-1}$ cm$^{-2}$, 
which is almost three times lower than the maximum flux detected 
with Ginga in 1987 (\cite{kit90}).
\pspicture(0,-1.8)(8.5,9.0)

\rput[tl]{0}(0,8.3){\epsfxsize=8.5cm
\epsffile[80 380 540 710]{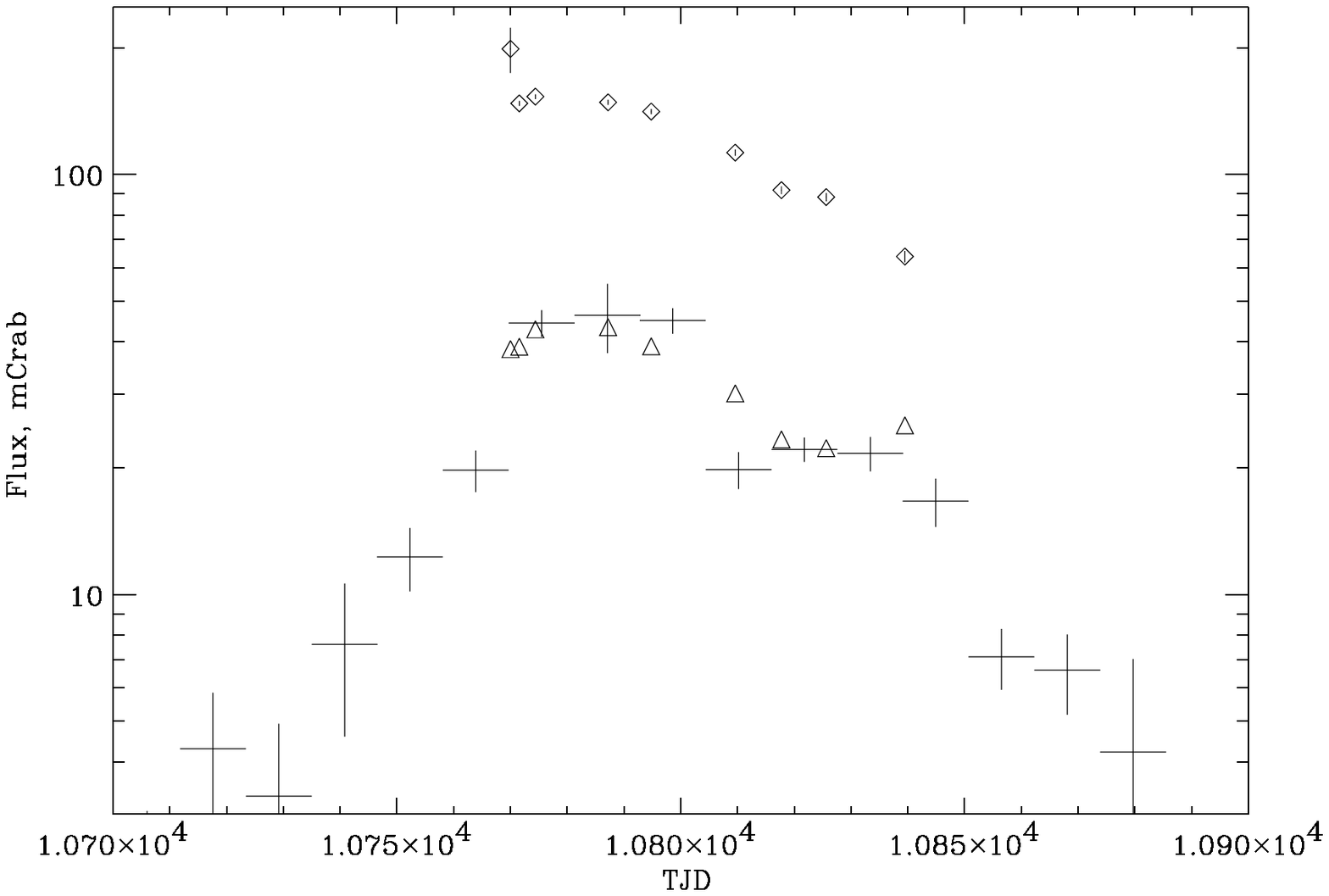}}

\rput[tl]{0}(-0.45,1){
\begin{minipage}{8.75cm}
\small\parindent=3.5mm
{\sc Fig.}~1.---Light curve for the 1997-1998 outburst of GS~1354--644. Time
axis in Truncated Julian Dates (TJD=JD-2440000.5). 
Crosses represent the ASM data (1.3--12.2 keV), 
open triangles - the PCA data (3--20 keV), and diamonds - 
the HEXTE data (20--100 keV), all normalized to the Crab nebula 
flux in the respective band.
\end{minipage}
}
\endpspicture

\subsection{Fast variability}

The PCA detected a strong flux variability as illustrated 
by Fig.~2.  The source flux varied by a factor
of 2-3 in 10 to 20 s. Such variability is often fit by
shot noise models (\cite{ter72}), in which the light
curve is made up of randomly occurring discrete and identical events,
the ``shots''. The approach was further developed (\cite{suth78}), and
has been proven to be quite useful in interpreting the archetypical
black hole candidate Cyg~X-1 (\cite{lochner91}) and other black hole 
binaries in their low state. We shall discuss below the application
of this model to GS~1354--644.

Another method to study fast variability is by means of Fourier
techniques (see the detailed discussion in \cite{klis89}), in particular,
by the analysis of the power density spectra (PDS). PDSs measured with 
the PCA for GS~1354--644 can be qualitatively described as
the sum of at least two band-limited components, each of which
is a constant below its break frequency, and 
a power-law with a slope $\sim$-2 above its break frequency.
PDSs for observations \#3 and \#9 are shown in Fig.~3.
The power spectra have been normalized as squared fractional $rms$,
according to Belloni \& Hasinger (1990), and rebinned logarithmically 
in frequency. Rudimental white-noise was subtracted. It was estimated
by Poissonian statistics as modified by dead-time effects.
We have plotted PDSs as $f \times \left({rms \over mean }\right)^2$ 
vs. frequency.  As was argued, e.g. by Belloni et al. (1997), this convention
has important advantages, namely, a plot gives a direct visual idea
of the power distribution, and Lorentzian functions representing 
band-limited noise are symmetric in a log-log plot.
This convention was used only for the plot, while all analytic fits
(see below) were performed on the original power spectra.
\pspicture(0,-1.0)(8.5,9.0)

\rput[tl]{0}(0,8.3){\epsfxsize=8.5cm
\epsffile[80 380 540 710]{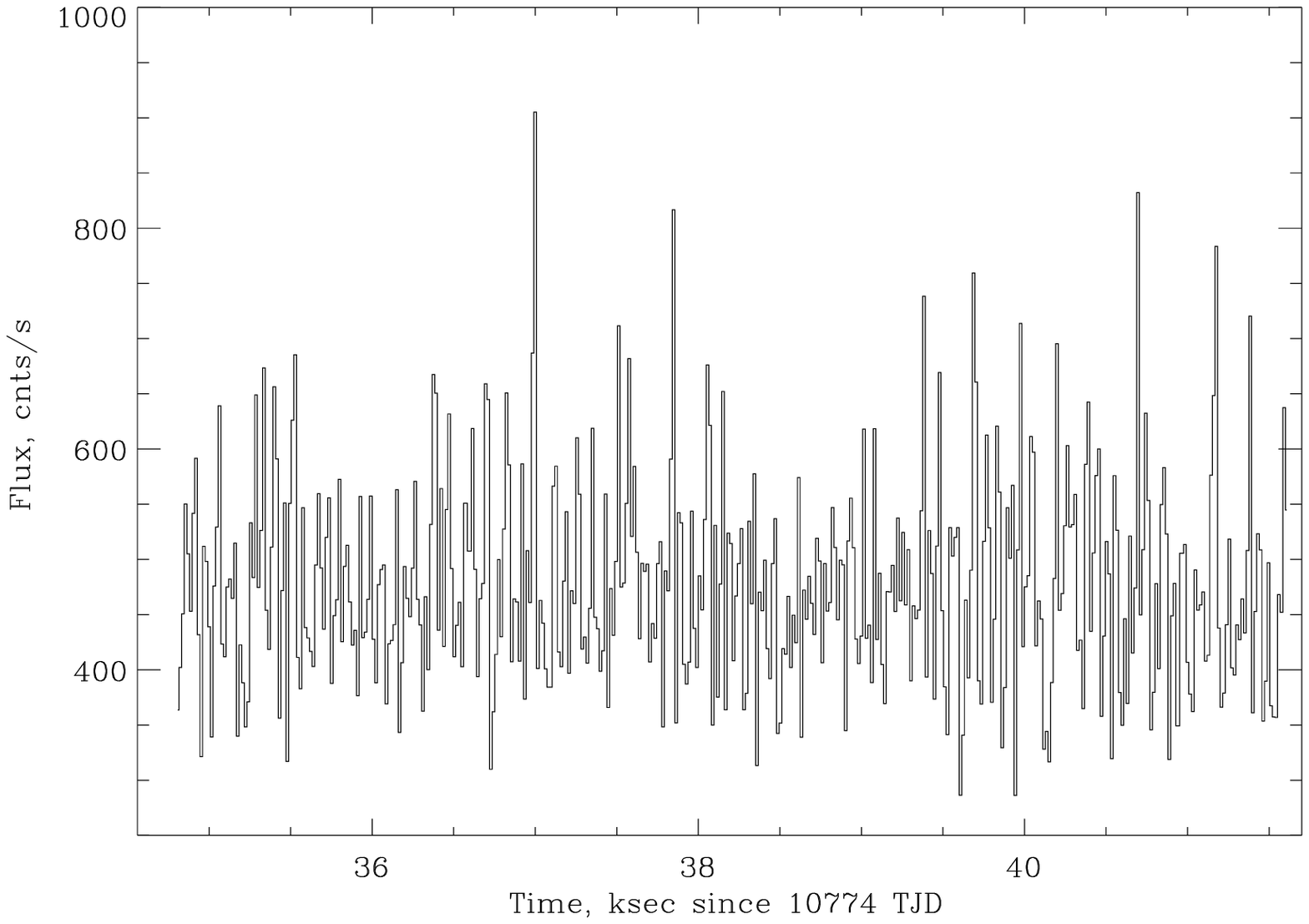}}

\rput[tl]{0}(-0.45,1){
\begin{minipage}{8.75cm}
\small\parindent=3.5mm
{\sc Fig.}~2.--- A typical time history of GS~1354--644, as observed by PCA. 
Each bin corresponds to a 16-sec time interval. 
Statistical errors are negligible on this scale.

\end{minipage}
}
\endpspicture

We fitted the PDS by a sum of functions 
$\sim {1\over{1+({f\over{f_{br}}})^2}}$, each of 
which represents the power spectrum from single type of exponential
shots with a profile $s(t) \sim exp({-(t-t_0)\over \tau})$,
for $t > t_0$, $\tau=1/(2\pi f_{br})$.  
The PDSs for observations \#5--9 are fitted satisfactorily
by two such components, but for the first four observations a third,
intermediate, component should be included for a good 
representation of the data. The fitting parameters are
presented on Table \ref{pds_par}. The inferred parameters
are very similar to those detected for other black holes
(e.g. \cite{nowak98a}; \cite{nowak98c}; \cite{grove}) and neutron stars
(e.g. \cite{oliv98a}) in their low state.  The frequency of the first break
in the PDS evolved significantly with time - it appeared 
at its lowest frequency near the outburst maximum, and shifted to higher 
frequencies with time. The frequency of the last break, however, 
remained fairly stable.

The energy-resolved power density analysis showed that 
the $f_{break}$ frequencies do not vary significantly with energy, 
similar to what is observed for Cyg~X-1
(e.g. \cite{nowak98c}) and other black hole candidates. 

The dependence of integrated fractional variability on energy in 
the full analyzed frequency range ($10^{-3}$--50 Hz) is presented in 
Fig.~4a (in $rms$ percent). 
The decline of integrated $rms$  with energy  
will be discussed in more detail in the next section, and
compared with data from other sources in section 5.3.

\subsection{Shot noise model}

Strong chaotic variability has been detected in many Galactic binaries
in their hard/low spectral state. 
Following the approach developed by Terrell (1972), 
such variability can be modeled in terms of a shot noise model 
(e.g. \cite{lochner91} and references therein). 
In the shot noise model, the light curve is 
assumed to be composed of a number of individual shots or microflares. 
In principle, different shots might influence each other, and might 
have various shapes and spectra.  However the models are usually
simplified to reduce the number of free parameters. We typically
assume that there are only a few types of shots, and that all shots
of a given type are identical. In this simplified model
statistical analyses of the observed light curve allow 
the determination of shot parameters.
For GS~1354--644, the overall shape of the power density 
spectrum suggests that the light curve is formed by two or
three different types of shots with characteristic times corresponding to
the breaks in PDS, $\tau=1/(2\pi f_{br})$(see Table \ref{pds_par}).
\pspicture(0,-2.5)(8.5,9.0)

\rput[tl]{0}(0,8.3){\epsfxsize=8.5cm
\epsffile[80 223 540 571]{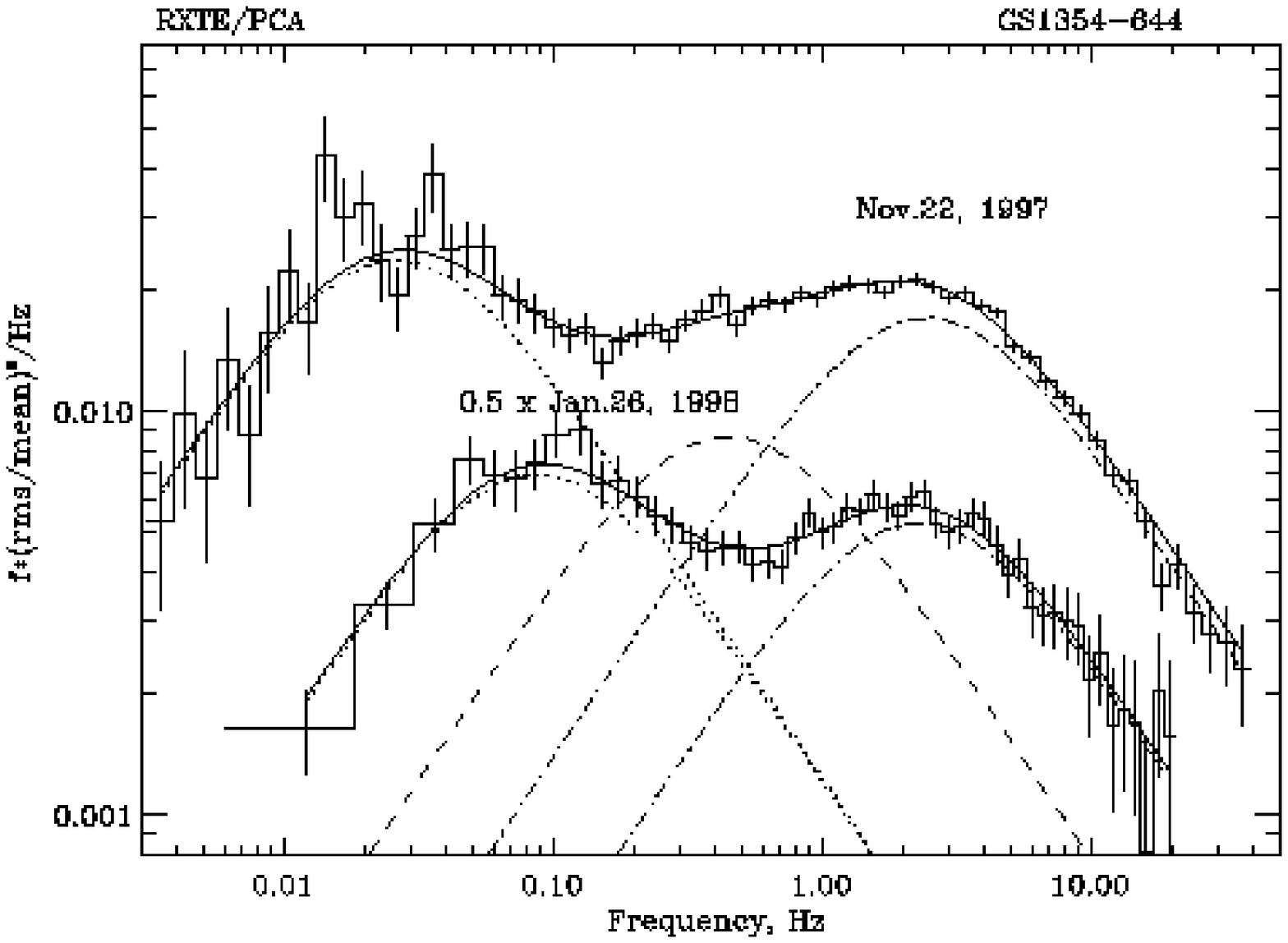}}

\rput[tl]{0}(-0.45,1){
\begin{minipage}{8.75cm}
\small\parindent=3.5mm
{\sc Fig.}~3.--- Power density spectra for two observations of
GS~1354--644 with  PCA. The upper spectrum was obtained for the
observation \#3;  lower one for \#9.
Three model components (see Table 2 and text) are shown by dotted,
dashed and dash-dotted lines respectively for both observations.
The data and model for observation \#9 (a two-component model was
applied in this case) were multiplied by 0.5 to avoid confusion with
observation \#3.
\end{minipage}
}
\endpspicture

The power density spectrum does not provide a complete 
description of the shots, because it is not possible to determine 
independently shot rates and their intensities.  To obtain additional
information we analyzed the flux histogram.
The probability histogram for the flux values
integrated into 16-sec bins is presented in Fig.~5.
We selected 16-sec time bins to study the long shots.  
The PDS shows that long shots and short shots
give almost equal input into the integrated fractional variability 
of the source,  but the contribution of short shots to the source 
variability is negligible below about 0.1 Hz.
The number of short shots in each 16-sec bin is
large enough that their contribution is nearly constant. 
To avoid the influence of Poissonian counting statistics on 
the distribution, we have chosen the flux bins 
to be 2 times wider than the Poissonian 
error associated with each bin.

For high shot rates one would expect that the distribution,
according to the central limit theorem, would have a symmetrical Gaussian
shape. However, this is not the case for our distribution, which has
a detectable deficit at lower fluxes.  This shape suggests
a low shot rate, and can be fitted by a Poissonian distribution,
if the duration of any single shot is substantially shorter than 
the 16-second bin width. 
For GS~1354--644 the first PDS component corresponds 
to time scales of 2-5 seconds, while 
the flux has been integrated into 16-sec bins.
We have assumed that the total flux from the source is 
a sum of a constant component, which is stable on
the time scale of one observation, and a variable component 
formed by individual shots. As was mentioned above, short shots 
appear as part of the constant component, so the flux variations
are caused by long shots only.
We have fitted the flux density distribution as a sum
of constant and variable components by applying 
the maximum likelihood method, which is preferable to 
chi-square statistics for low event rates.

\pspicture(0,-2.5)(8.5,13.7)

\rput[tl]{0}(0,13){\epsfxsize=8.5cm
\epsffile[54 361 590 720]{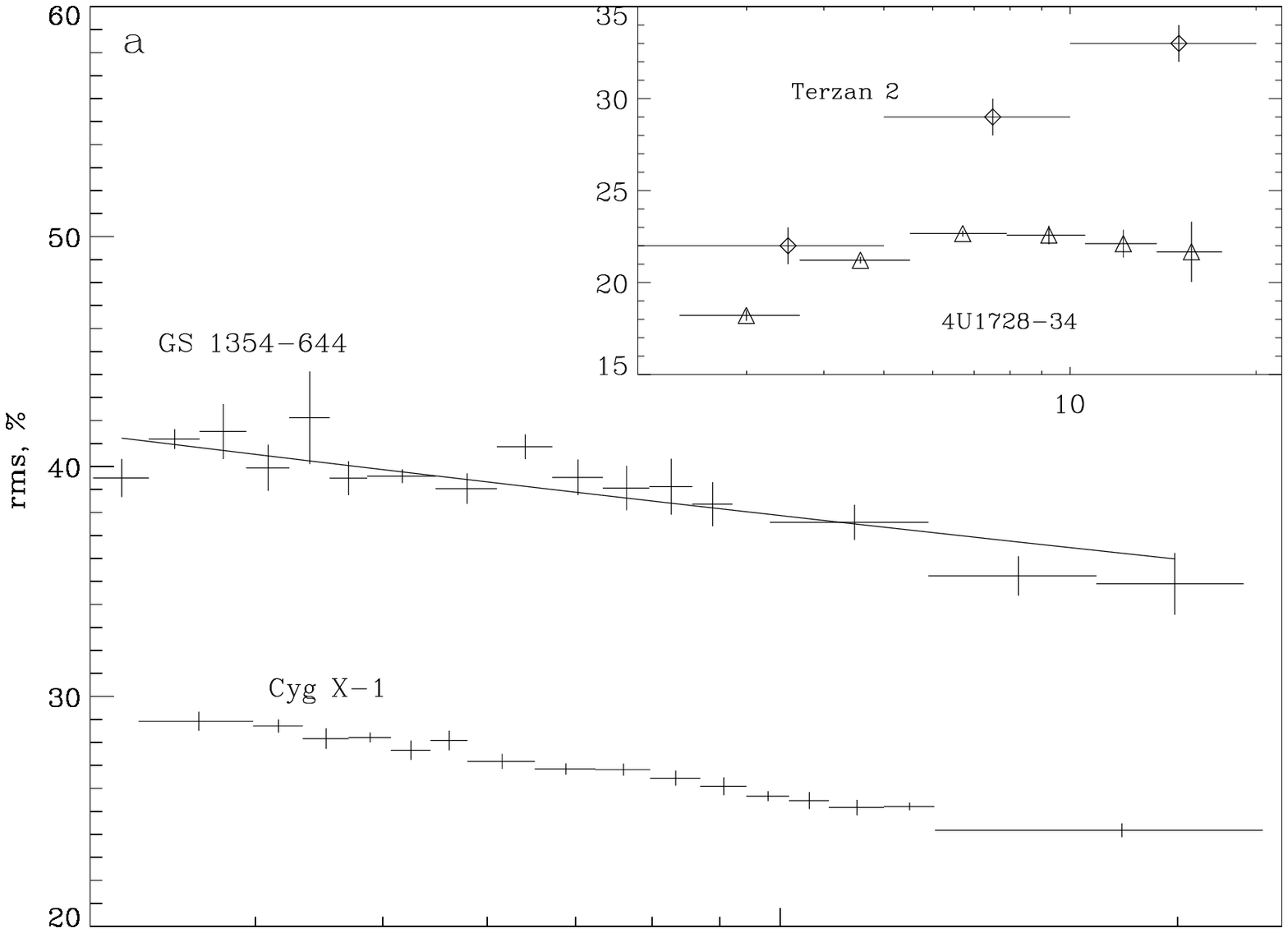}}

\rput[tl]{0}(0,7.3){\epsfxsize=8.5cm
\epsffile[54 361 590 720]{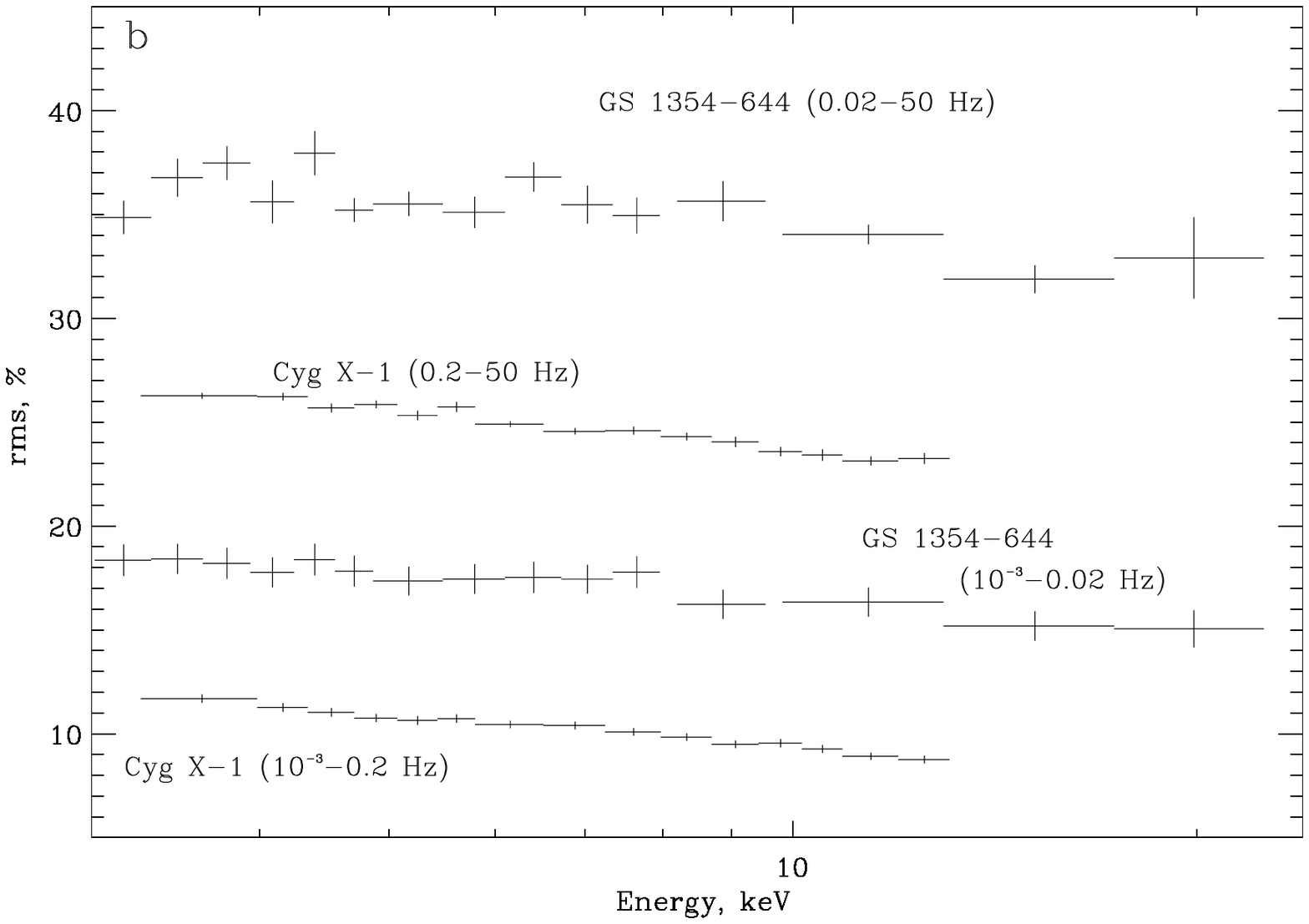}}

\rput[tl]{0}(-0.45,1){
\begin{minipage}{9.0cm}
\small\parindent=3.5mm
{\sc Fig.}~4.--- 
{\it Upper panel(a)}: The dependence of the fractional variability 
(\% $rms$) of GS~1354--644 flux integrated over the frequency range 
$10^{-3}$--40 Hz on photon energy (observation \#3). Dependences 
for Cyg~X-1, 4U1728-34 and Terzan~2 (Olive et al., 1998a) 
are also shown for comparison.
{\it Lower panel(b)}: Energy dependence of the fractional variability
(\% $rms$) evaluated over different frequency ranges (above and below 
the first break in the PDS) for GS~1354--644 and Cyg~X-1.
\end{minipage}
}
\endpspicture

The best fit for the shot rate was $\simeq0.3$ shots per second, 
and the best fit for constant component was in the range of 50--70\%
of the total flux (in reality this method only puts an
upper limit of the value of the constant component. Our analysis of
0.01-0.1 sec light curve showed that there exist points with the flux
$\sim$10 times lower than the average source flux. This fact
immidiately removes the upper limit of the constant component down to
the 10\% of the average flux). 
The signal detected from an individual shot was estimated to be 
$\sim$500 PCA counts for the first observation of 1997, and
$\sim$200 PCA counts for the last outburst observation (\#9). 
The best fit obtained for observation \#4 is presented in Fig.~5. 
We repeated the analysis for a 64-sec
integration time and obtained consistent results.
The low shot rate for long shots shows that the variability 
of the source on time scales of tens of seconds is caused by
relatively rare powerful flares.

\pspicture(0,-1.5)(8.5,9.0)

\rput[tl]{0}(0,8.3){\epsfxsize=8.5cm
\epsffile[80 220 570 570]{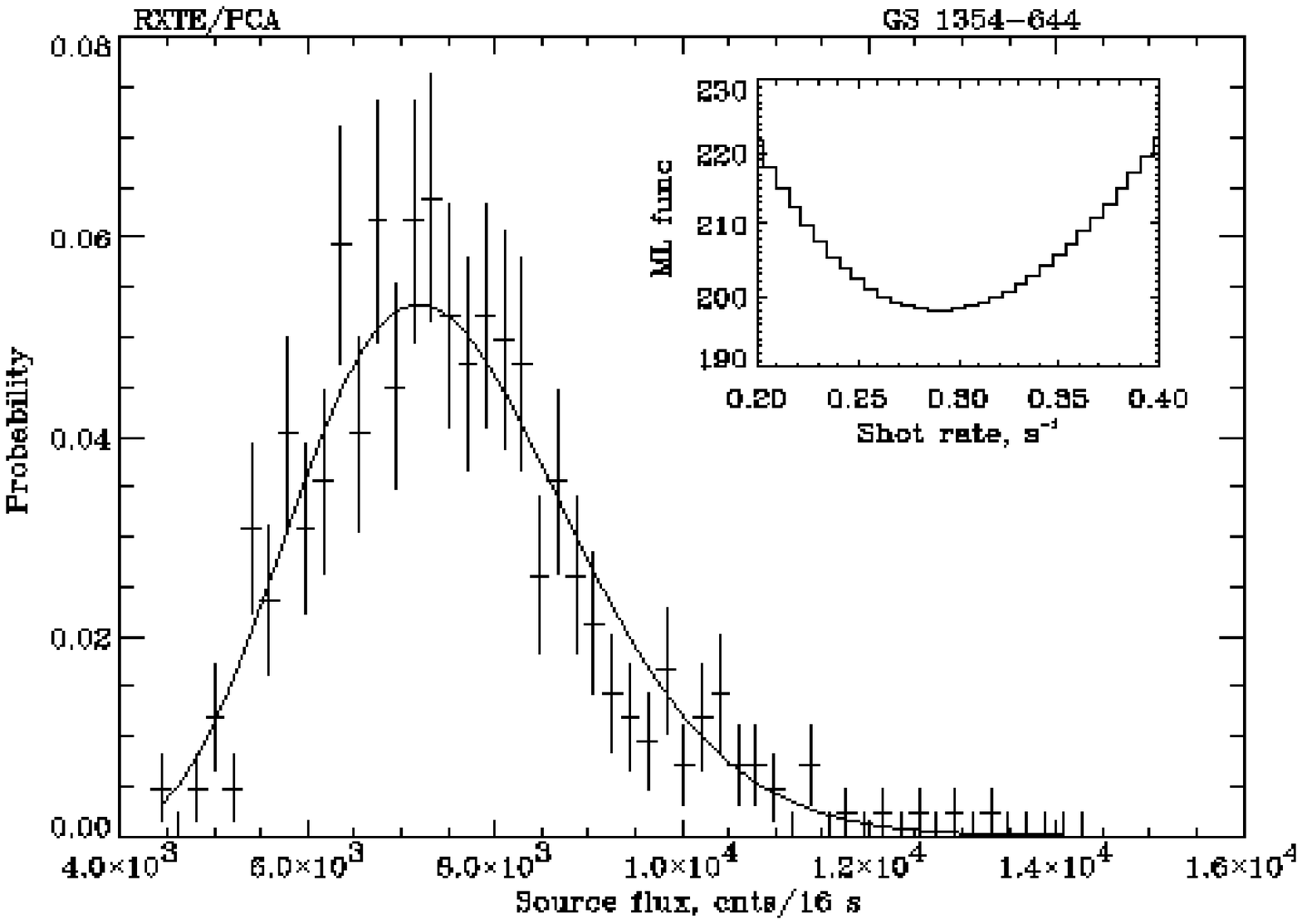}}

\rput[tl]{0}(-0.45,1){
\begin{minipage}{8.75cm}
\small\parindent=3.5mm
{\sc Fig.}~5.--- The distribution of the source flux integrated in 16-sec 
bins (observation \#3).
The dependence of the maximum likelihood function on shot rate is shown in
the upper right corner. The best fit parameters are: 
a single shot amplitude $\sim215$ counts and a shot rate of
$\sim$ 0.3 shots/s.
\end{minipage}
}
\endpspicture

Unfortunately, we were not able to study the contribution of short
shots by the same method, because the flux distribution for 
time scales of tenths of a second depends strongly on both the short 
and long shots. However, the integrated $rms$ variability of
the short shots, combined with some simple constraints on 
the shot amplitude, allows us to estimate the shot overlapping 
parameter (see \cite{vihl95} for the method description), 
and consequently derive the shot rate. The short shot rate 
can be estimated to be $\sim$10-15 shots/s.

The physical origin of the shots is not very clear. For GS~1354--644, 
as well as for other black hole systems in their low state,
the break frequencies detected in PDS ($\sim$1 Hz) are much lower than 
would be expected for the environment close to a stellar 
mass black hole ($\sim10^2$ Hz). The shape of the energy spectrum, 
which is dominated by the Comptonized emission component, moves us to
explore whether the Comptonization might be responsible 
for the time blurring of intrinsically short shots.
(This could be the case if the product 
of the Compton optical depth and  
light crossing time is of order one second).
The detected dependence of fractional variability on energy 
(Fig.~4a) at first glance 
seems to support this interpretation, because photons of higher 
energy undergo on average more interactions and so must have a wider
distribution in time and lower integral variability.  However,
the fractional variability integrated up to the lowest break in PDS
should not be affected by this mechanism and should therefore 
be independent of the energy.  We tried to test this assumption 
for GS~1354--644, but found that slope of the dependence of $rms$ 
variation vs energy for the frequencies below 0.02 Hz 
(see Fig.~4b)
cannot be defined accurately. To get a more definitive answer 
we had to repeat our analysis for a much brighter source, 
Cyg~X--1, which in many ways resembles GS~1354--644.  
For Cyg~X--1 we can clearly see that the fractional variability 
decreases with energy both when integrated up to 50 Hz, 
and when integrated for the frequency range below 0.2 Hz.  
This last result is in direct contradiction with an assumption that
the Compton up-scattering is responsible for the low-state PDS break
frequencies.

\subsection{Time lags in GS~1354-644}

\pspicture(0,-0.5)(8.5,9.0)

\rput[tl]{0}(0,8.3){\epsfxsize=8.5cm
\epsffile[49 190 570 590]{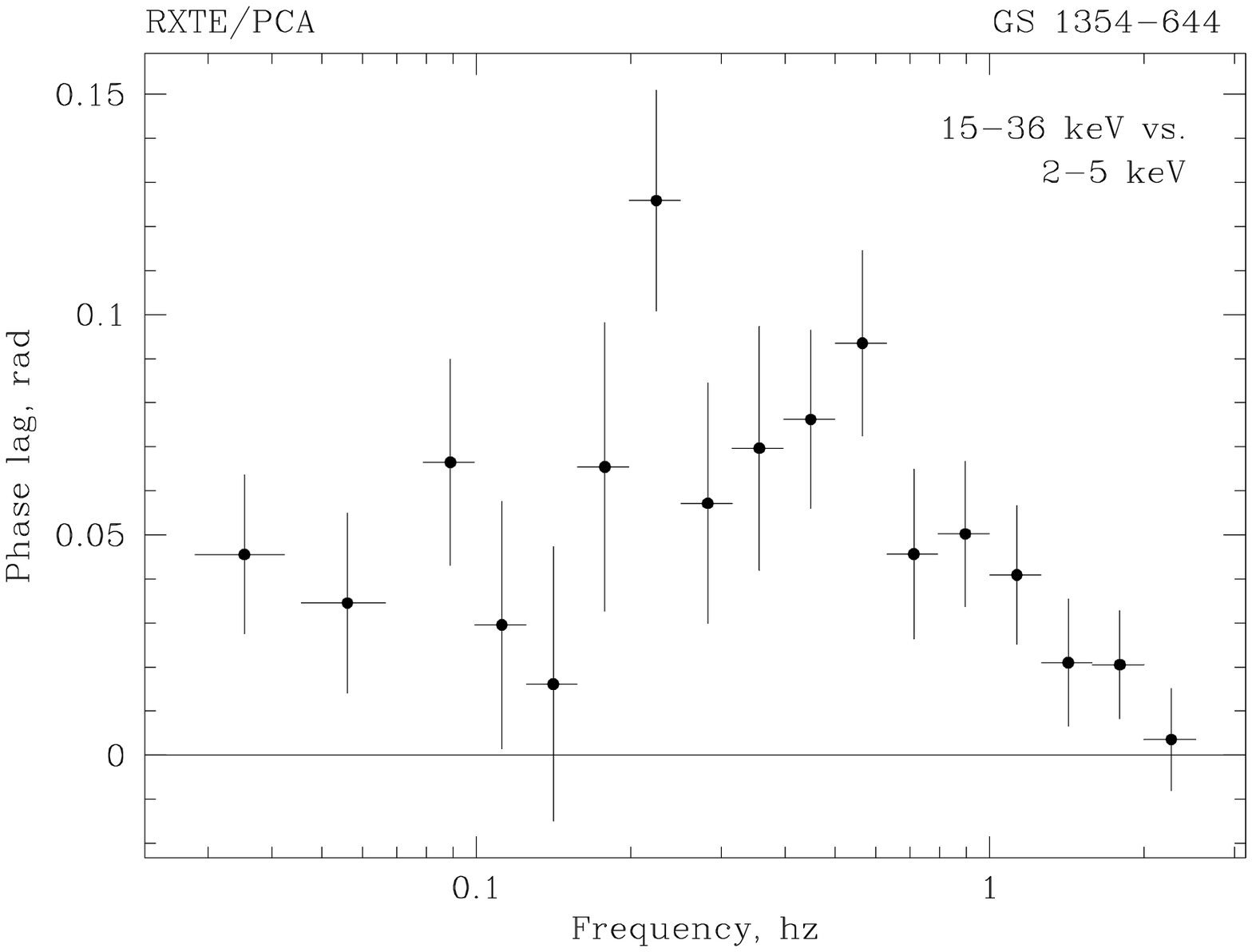}}

\rput[tl]{0}(-0.45,1){
\begin{minipage}{8.75cm}
\small\parindent=3.5mm
{\sc Fig.}~6.--- Phase lag as a function of Fourier frequency for the 15--36
keV energy band vs. the 2--5 keV energy band.
\end{minipage}
}
\endpspicture

Another way to study rapid fluctuations in the flux is to compute
time lags between variations in different energy bands.
We calculated frequency-dependent phase lags according to the
procedure of Nowak et al. (1999a). Due to the faintness
of the source we were forced to sum almost all of the data available 
-- from observations \#2 through \#9 -- to obtain a significant
result. The total integrated live time for these PCA observations 
was $\sim48$ ksec.  Time lags are presented in Fig. ~6.
The error values were estimated from the width of the
lag distribution histogram. One can see that, at least
qualitatively, the dependence of phase lag on Fourier frequency
is very similar to low states in other X-ray binaries, both in 
black holes and neutron stars  (e.g. Nowak et al. 1999a, 
Grove et al. 1999, Ford et al. 1999).

\section{Energy spectrum}

The energy spectrum of GS~1354--644 during its 1997 outburst can be 
roughly represented as a hard power law with slope $\alpha \sim$1.5, 
and a high-energy cutoff above $\sim$50 keV. Such spectra are 
typical for black hole binaries in the low/hard spectral state.  
A commonly accepted mechanism for generating the 
hard radiation in this state is thermal Comptonization  
(\cite{sle76}; \cite{st80}).  Applying a thermal Comptonization
model to the hard-state BH spectra, one infers a hot
cloud surrounding the central object (BH) with typical plasma
temperature $\sim$50 keV and a Thomson optical depth $\ga$1. 
Such a cloud can Comptonize the soft photons from the
central region, most likely from an accretion disk.

\pspicture(0,-2,5)(8.5,9.0)

\rput[tl]{0}(0,8.3){\epsfxsize=8.5cm
\epsffile[36 185 580 590]{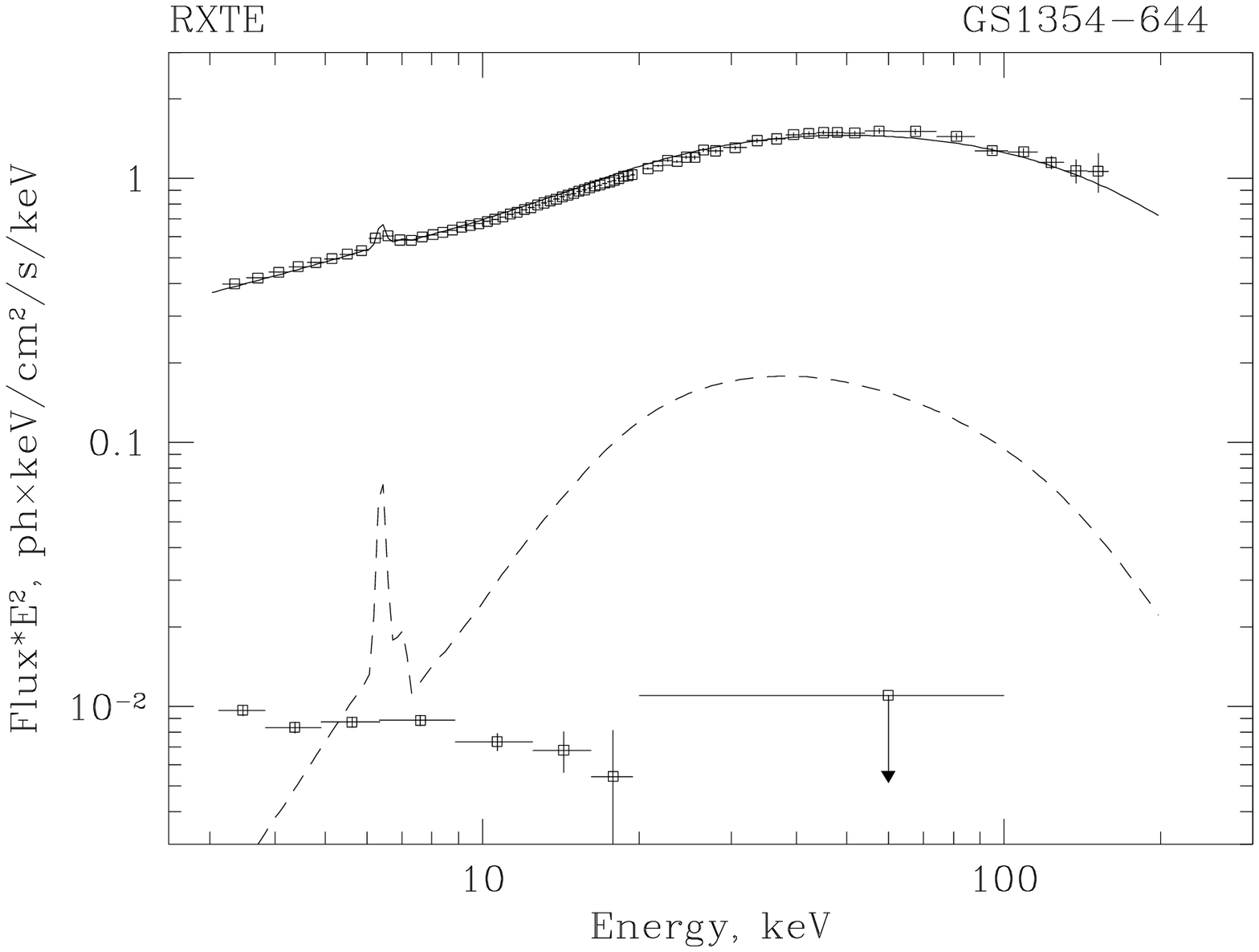}}

\rput[tl]{0}(-0.45,1){
\begin{minipage}{8.75cm}
\small\parindent=3.5mm
{\sc Fig.}~7.--- Spectral energy distribution of GS~1354-644 during the 
outburst maximum and in quiescence. The upper spectrum is an average 
over observations 2--5, and the lower one is from  
observation \#10 (there was no detection by HEXTE, 
so a 2$\sigma$ upper limit is shown for energies above 20 keV).
The solid line is a spectral fit to the data by the model 
described in the text ($compST+reflection+Gauss$; see Table
\ref{spectra_par}).  The components produced by adjacent cold matter
through Compton reflection and Fe $K_\alpha$ fluorescence are shown
by the dashed curve. 
\end{minipage}
}
\endpspicture

We fitted the spectrum with a variety of models, ranging from a simple
power law to detailed Comptonization, in search of the underlying
physics.  The results are presented in Tables \ref{spectra_par} and
\ref{spectra_hexte_par}.
We present results for two groups of observations (2-5 and 6-9),
because the spectral parameters did not differ significantly
within each group, but summing the groups allows a more accurate 
parameter estimates.  The spectra for observations 2-5 
and also for observation \#10 are presented in Fig.~7.
The y-axis is in units of the photon flux multiplied by the energy squared.
These units show directly the energy content per decade.
Crab Nebula spectrum plotted the same way would be close to flat,
and Galactic black hole binaries typically show a negative slope
in their high state, and a positive slope (up to $\sim$100 keV) 
in their low state.

On using the same models for the PCA and HEXTE data considered
jointly we found that cross-calibration uncertainties 
of these instruments played a significant role. 
Namely, any power law (with high energy cutoff) approximations
gave noticeably different photon index values, even if the same energy
bands were used.  So we used power laws with photon indexes 
differing by 0.08-0.1 for PCA and HEXTE spectra (we followed here
an approach by \cite{wilms}). In the Table \ref{spectra_par} we 
present the PCA photon indices.    

The fits show a noticeable high energy cut-off at energies above 
$\sim$60 keV. While different models give different cut-off 
parameters, no fit without a cut-off was satisfactory.
In fact, the spectrum cannot be fully described by a simple power law, 
with or without a high energy cut-off. The spectral fit is improved 
significantly by adding a neutral iron fluorescent line 
and a reflection component, which indicates that 
some part of the emission is reflected by optically thick cold 
material, most probably in the outer accretion disk 
(\cite{basko}; \cite{gf91}; \cite{mz95}).  
This component was represented by $pexrav$ model
of $XSPEC$ package, which simulated reflection from a neutral medium.

For Comptonization models, we applied the classical $compST$ model 
of \cite{st80}, and the more recent generalized $compTT$ model (\cite{tit94}).
For both cases the parameter $E_{cut}$, the cutoff energy for the
$pexrav$ model (reflected component), 
has been frozen at the value $3kT_e$, where $kT_e$ is the temperature 
of the Comptonizing hot electrons. The value of the optical depth 
parameter depends on the assumed geometry - for $compTT$ we cite 
the $\tau$ parameter for both spherical and disk geometries. 
It is well known that the cut-off at energies higher than $\sim3kT_e$ 
is more abrupt in the spectrum of Comptonized emission than 
a simple exponential cut-off, but we were not able to detect 
a statistically significant difference between these two cut-off 
models because of the faintness of the source. A Gaussian component 
with the central energy 6.4 keV and FWHM equal to 0.1 keV 
(frozen at these values) was added to account for emission 
at the iron line. 

During the observation of 17 November 1998 (\# 10) the PCA
measured a very low flux.  The spectrum was found to be softer, 
with a power law photon index $\sim$2 in comparison 
with $\sim$1.5 the year before.
This observation might represent the quiescent state of GS~1354--644 
or the transition to quiescence.

\pspicture(0,-1.5)(8.5,9.0)

\rput[tl]{0}(0,8.3){\epsfxsize=8.5cm
\epsffile[80 380 560 710]{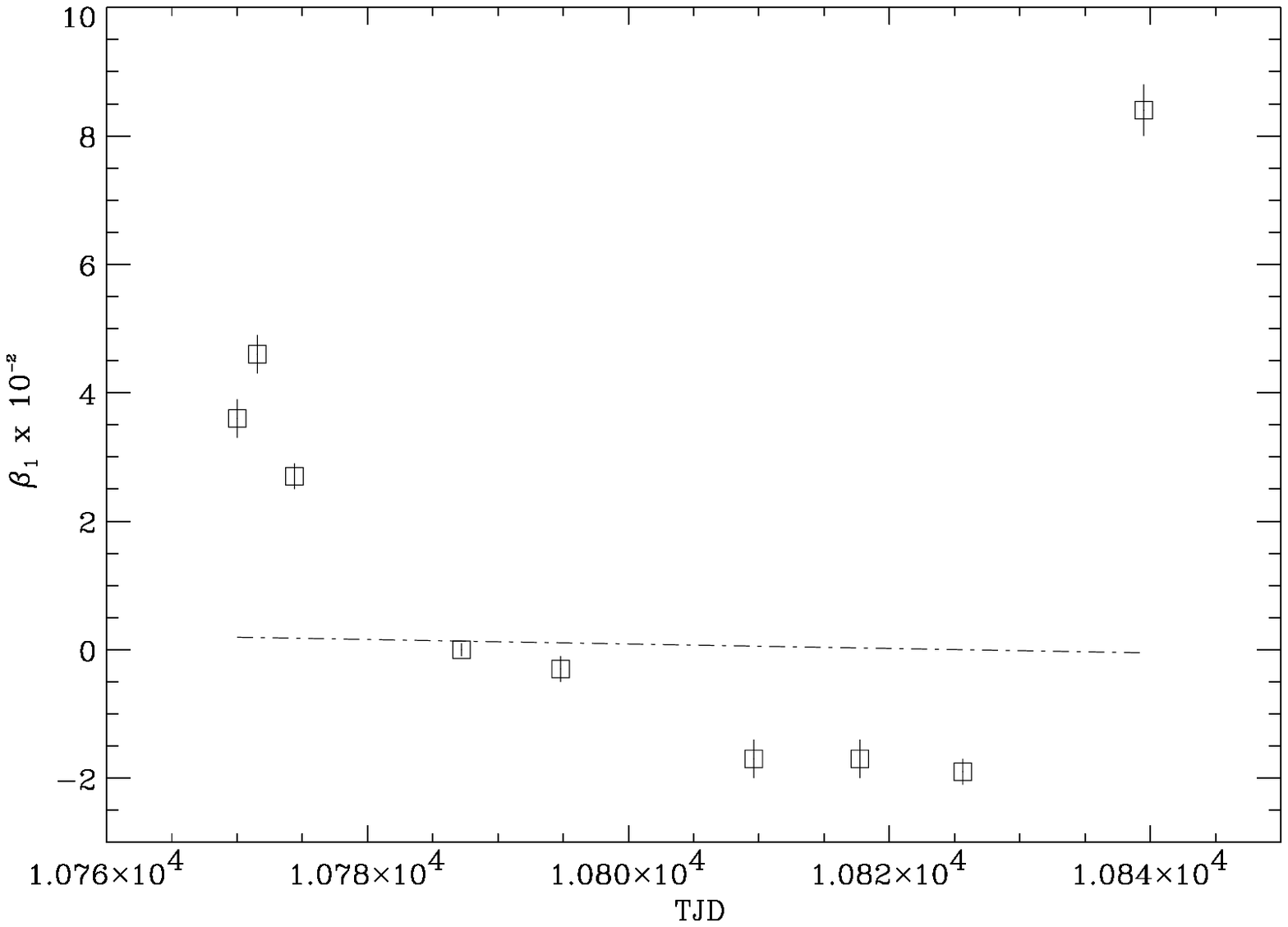}}

\rput[tl]{0}(-0.45,1){
\begin{minipage}{8.75cm}
\small\parindent=3.5mm
{\sc Fig.}~8.--- The evolution of the spectral ratio slope (see text) 
during the transient outburst. 
Observation \#4 was used as a reference spectrum. 
The dot-dashed line shows the instrumental stability obtained from
observations  of the Crab nebula. We note an overall softening 
of the spectrum through observation \#8.  
Observation \#10, made a year later, is not shown.
\end{minipage}
}
\endpspicture

\subsection{Spectral change analysis}

To study spectral changes in more detail, we analyzed raw spectral
ratios.  This method is based on the idea that, if one divides one
spectrum by another, subtle differences between two are
evident in the ratio.  We applied this method to spectra obtained
during different PCA observations and also to spectra collected
within each observation, segregated by flux level.  
In all cases, the spectral ratios could be approximated 
by a single power law $E^{\beta}$ in spite of the complexity 
of the initial spectra.  We have used the power law index $\beta$
for quantitative comparisons of spectral ratios.  We labeled 
this index as $\beta_1$ when the method was applied to different 
observations, and $\beta_2$ when we compared high-flux and low-flux 
spectra for the same observation.  In both cases we call $\beta$ 
spectral ratio slope or differential slope.

For the separate observations, we divided each spectrum by the spectrum
of the observation with maximum flux (\#4).  This method allowed us 
to exclude from the analysis uncertainties in the PCA response 
matrix, and hence to study fine differences between spectra.
The drift of detector parameters, however, can contribute a significant
systematic error, if the interval between observations is too long.
To estimate the level of such systematic errors, 
we applied the same technique to Crab spectra taken before, 
during and after the outburst of GS~1354--644.  The apparent 
changes in the Crab spectral ratio caused by drifting detector response
are much smaller than the changes seen in GS~1354--644 
(Fig.~8).  The X-ray spectrum of GS~1354--644
became softer with time, then suddenly harder for observation \#9.
There is no clear correlation with flux level.  
Instead, observations at the same flux level taken during rise
and decline had significantly different spectral ratios
(see differential slopes $\beta_1$ on Table \ref{spectratio_tbl}).

\pspicture(0,-1.3)(8.5,9.0)

\rput[tl]{0}(0,8.3){\epsfxsize=8.5cm
\epsffile[80 380 560 710]{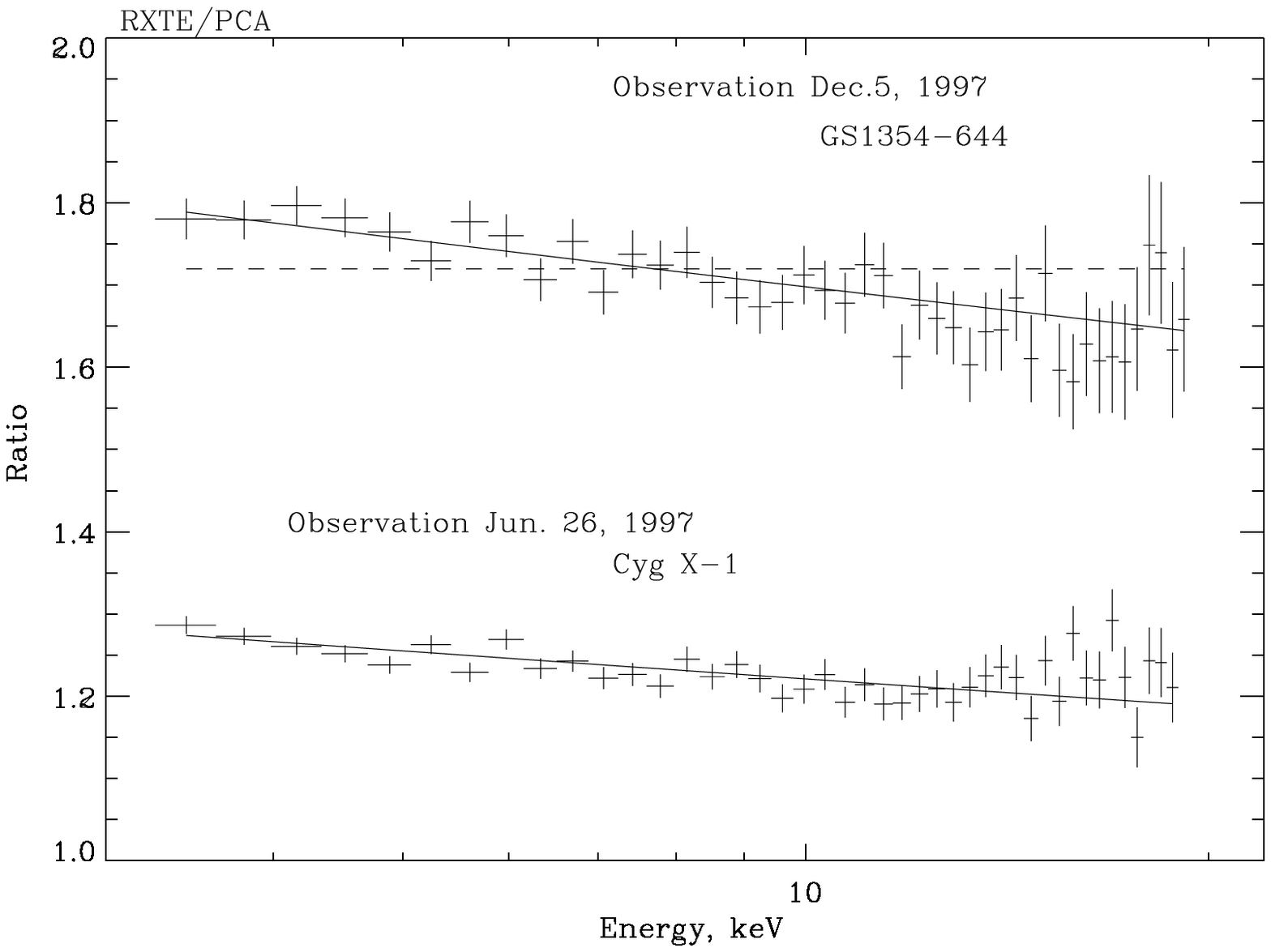}}

\rput[tl]{0}(-0.45,1){
\begin{minipage}{8.75cm}
\small\parindent=3.5mm
{\sc Fig.}~9.--- The 3-20 keV spectral ratio for GS~1354--644
(observation \#4)  obtained by segregating higher and lower flux
intervals, as described  in the text. A similar ratio for Cyg~X-1 is
shown for comparison.  In both cases the spectrum softens at higher
flux levels. 
\end{minipage}
}
\endpspicture

The same technique was applied to study the relation between spectrum
and flux within each observation.  In this case
the data were segregated according to the total count rate 
averaged into 16-sec bins.  The range from minimum to maximum flux
was divided into three equal parts, and two spectra for ``high''
and ``low'' fluxes were obtained. 
Thanks to the large chaotic variations in the flux, the ``high'' 
and ``low'' average levels differed by a factor of 1.5--2. 
Finally, for each observation the high-flux spectrum was divided
by the low-flux spectrum to obtain a spectral ratio. 
A typical high-flux/low-flux spectral ratio is shown in 
Fig. ~9 (from observation \#4).
A negative differential slope was obtained for all other observations 
within the outburst. It is evident from Table \ref{spectratio_tbl} 
that while spectral ratio slopes $\beta_1$ for separate sessions are not
correlated with flux, high-to-low spectral ratios of a single
observation always have slopes $\beta_2 <$ 0.
For comparison, we present in Fig. ~9
the spectral ratio for Cyg~X-1 calculated with the same technique.  
The effect is qualitatively the same for Cyg~X-1 as for GS~1354--644,
with a differential slope $\beta_2 = (-3.9 \pm 0.4) \times 10^{-2}$
(observation 26 June 1997).

We were concerned that selection effects might bias our estimates
of $\beta_2$.  Because most counts fall in the soft part of 
the spectrum, one might get softer spectra simply by
selecting for high flux, because we would be selecting for 
random excesses in the soft band. 
To check the importance of this effect, we changed our selection 
criteria and re-filtered the data according to the flux levels 
above 10 keV. The results remained the same, indicating that 
the observed correlation of higher fluxes with softer spectrum  
was much stronger than any bias introduced by the selection criteria.

The systematic difference between high and low flux spectra 
can be interpreted in a variety of ways.
The softening may be due to an increase in the relative contribution 
of shots that have a softer intrinsic spectrum than the constant flux. 
The constant flux, in turn, may be formed by the sum of
shorter shots with harder spectra (this possibility is discussed, e.g.
by \cite{freq_resolved_rev} and \cite{tim_cygx1}). Another interpretation might be that 
the low energy shot emission changes the temperature of the hot plasma 
cloud responsible for Comptonization. 
Alternatively it might be suggested that the shot spectrum evolves 
with time and is softest near the maximum of the shot.
Other explanations can not be excluded either.

Our spectral ratio analysis is quite sensitive
to subtle changes in parameters of the system. 
We see that, in the course of the 1997 outburst, the spectrum
softens steadily (except for observation \#9).  This implies
that, in this case, the spectrum was not directly correlated with 
the flux level (or luminosity). We detected a hardness-flux 
anti-correlation for short-term flux variations within each 
observation, but not when we compared spectra for different 
observations. The last observation of the 1997 outburst, 
which breaks the overall pattern, likely corresponds to the secondary 
maximum or 'kick' in the light curve. The observation \#10,
performed a year after the maximum of 1997 outburst, showed 
a significantly softer spectrum with flux more than two orders 
of magnitude below the peak value.

\section{Discussion}

\subsection{Spectral state}

The high and low spectral states first identified in Cyg~X-1 
(\cite{tan72}) have since been observed in a number of X-ray binaries.  
The ``high'' and ``low'' terminology was originally chosen based on
the 2-10 keV X-ray flux. It was later found that the {\it low} state
corresponds to a hard spectrum, and the {\it high} state to a soft 
spectrum. A typical {\it high} state spectrum is the sum of
a soft thermal component and a hard power-law tail.  The
{\it low} state spectrum is approximately power-law 
with an exponential cut-off at energies above $\sim$ 100 keV. 
More detailed descriptions of the spectral
states of X-ray binaries can be found elsewhere (e.g. \cite{tsh96}).

Studies of the aperiodic time variability in black hole candidates 
provided another dimension to this phenomenology. In particular,
a third very-high state has been recognized (Miyamoto et al. 1991, 1994;
\cite{ebi94}). This state is characterized by a 3-10 Hz QPO peak, 
plus either band-limited noise or a weaker power law noise component
(\cite{bel97}).

The hard power-law-like energy spectrum with a slope $\sim$1.5
that was detected from GS~1354--644 during its 1997 outburst
is a clear indication of the low/hard spectral state.
The character of the rapid variability, the absence of QPO, and 
the power spectrum of GS~1354--644 are similar to 
other black hole candidates in this state
(\cite{miya92}; \cite{nowak98a}; \cite{nowak98c}; \cite{grove}), 
and provide additional proof for such characterization. 
A similar time variability is manifested by neutron star binaries 
in their low state (\cite{oliv98a}; \cite{ford98}).  
However, black hole systems in their other - high or very high - 
states have distinctly different power spectra 
(\cite{klis95}; \cite{bel97}).

GS~1354--644 was detected with Ginga in 1987 in its high/soft state 
(\cite{kit90}).  But in 1997-1998 the same source 
was in a low/hard state.  This confirms the earlier identification of 
the source as a black hole candidate and also demonstrates that both
high/soft and low/hard states are generic for this group of X-ray 
sources. In fact, all the properties of GS~1354--644, that we revealed 
in our analysis are similar to other black hole candidates, 
both persistent and transient.

\subsection{Geometry of the system}

The energy spectrum of GS~1354--644 supports a model of
low energy photons that are Comptonized in a hot plasma cloud. 
An accretion disk reveals itself via fluorescent
iron line emission detected at $\sim$6.4 keV and a reflected continuum 
detected at 15-20 keV.  The equivalent line width 
and relative intensity of the reflected component are both 
consistent with reflection of harder X-ray emission from a cold plasma, 
which, in the commonly accepted model, is likely to be an optically 
thick accretion disk around the compact object.
It is straightforward to assume that the X-ray source 
includes a cool disk plus a hot optically thick corona, which 
Comptonize the soft photons up to tens of keV.
The exact geometrical configuration of the system remains uncertain. 

This disk/corona combination is the most popular model for
interpreting Galactic black holes in their low/hard spectral state.
A slab geometry, where the accretion disk is sandwiched
between two flat corona layers, was widely discussed 
several years ago (\cite{hm91}, \cite{har93}), but cannot fit
fit the spectra in a self-consistent manner (see eg. \cite{dove97}).
Other investigators have suggested a spherical hot corona (\cite{kht97}) 
or an advection-dominated accretion flow (\cite{nar96}) 
at radii smaller than the inner edge of the thick accretion disk.  
Seed soft photons could be generated in the accretion 
disk (\cite{dove97}, \cite{dove98}) or inside the hot cloud
(suggested by \cite{nar96} to be synchrotron/cyclotron photons).
The former models invoke external illumination of the up-scattering
region, while the latter invoke a source embedded inside
the hot plasma.

The optical depth $\tau \sim 4-5$ inferred from our spectral
fits (for $compST$ and $compTT$ models in a spherical geometry) 
is dependent both on the geometry and on the model, and hence 
can not be considered as a direct measurement of the optical depth.  
However, the value of $\tau$ indicates that the source 
of seed photons is most likely {\it inside} the hot cloud. 
In the case of external illumination of the Comptonization region 
by soft photons originating from the accretion disk, 
the optical depth parameter would be expected to be $\sim$1 
even for high intrinsic values of $\tau$ in hot cloud.
A partial overlap of the disk by the corona may solve the puzzle, 
but the physical basis for such a configuration has yet to be explored.

If the characteristic frequency of the Compton cloud is 
the PDS break frequency about $2-3$ Hz, then the cloud should be huge, 
which causes severe problems with energy balance 
(see e.g. \cite{nowak98a}).  More likely, the breaks in the PDS 
are determined by the intrinsic duration of the seed shots.
They might also be related to the geometrical parameters of the system,
such as the radius of the accretion disk, and with typical times,
such as the plasma diffusion time.
In the PDS of GS~1354--644 the first break frequency is 
anti-correlated with the flux, perhaps reflecting changes 
in some of the system parameters.
For example, the inner edge of the disk is probably moving closer 
to the compact object when the mass accretion rate grows 
and the luminosity increases.  Simultaneously the break frequency
is decreasing, which might mean that shots become longer.

\subsection{Dependence of fractional variability on energy}

Fig. ~4 shows that the integrated fractional 
variability of GS~1354--644 clearly decreases with energy.
This decrease can be approximated by a power law $rms\sim E^{-0.07}$.
Similar dependences have been found for 
Nova Per 1992 (\cite{vihl95}), Cyg~X-1(\cite{nowak98a}), and
GX~339--4(\cite{nowak98c}), all Galactic black hole binaries, when 
observed in the low/hard state.
In contrast, an {\it increase} of fractional variability with energy
has been detected for X-ray bursters -- e.g. 1E1724-3045 
(\cite{oliv98a}), 4U1608-522 (\cite{yu}) in their low/hard state.  
This apparent difference between low-state
black hole and neutron star systems 
is remarkable, because these systems are otherwise so similar, 
when in their low/hard state.
This fact motivated us to perform our own survey of
publicly available RXTE data.  
The X-ray bursters 4U1705-44, SLX~1735-269, 
SAX~J1808-3659, 4U1728-34 (GX~354-0) and 4U0614+091, 
together with the aforementioned
systems 1E1724-3045 and 4U1608-522, were included in our
analysis. 

Power-law fits to the relation between $rms$ variation and energy
are presented on Table \ref{rms_en}. All observations 
were taken when objects were in their low/hard state. 
For GX~339-4 and Terzan~2, the data
were taken from \cite{nowak98c} and \cite{oliv98a} respectively.
Although a simple power law was in some cases a poor fit,
we used it anyway to quantify the tendency (whether the $rms$ 
variation decreased, increased or stayed constant with energy).
Remarkably, slope is negative for all black hole binaries
and positive for all neutron star binaries. The strongest correlation
of fractional variability with energy is at energies below $\sim$15~keV.  
In fact, the fractional variability
has a broad maximum at energies 10-20 keV for many X-ray bursters, 
whereas for black holes the maximum is at the lowest observed
energy.  This might indicate that the
similar spectra in black hole and neutron star systems 
have fundamentally different origins.

Whatever the physical reason, the difference is quite remarkable 
because of the substantial similarity in the other properties 
of black hole and neutron star systems in the low/hard state 
(e.g. \cite{berg98}; \cite{revnivtsev_asca}; \cite{oliv98a}).  
If confirmed, this dependence may become an important new tool 
for deriving the type of compact object in Galactic binaries 
from X-ray observations.

\section{Conclusions}

We analyzed observations of the recurrent X-ray transient GS~1354-644
by the RXTE satellite.
The observations were made during a modest outburst of the source
in 1997-1998. The overall light curve was triangular 
with possible plateau at the maximum. 
PCA/HEXTE observations were carried out during both
the rise and the decay phase of the outburst.  
The dramatic fast variability 
was studied in terms of a shot noise model.
The power density spectrum can be approximated by the sum of 2 or 
3 components, each corresponding to a specific type of shot.
The most prominent components peak around 0.02--0.09 Hz 
and 2.3--2.9 Hz respectively.  For several observations 
a third, intermediate component is statistically significant.  
Our flux density distribution analysis showed that the longest
shots occur at a rate of 0.3 $s^{-1}$, and contribute 30--50\% 
of the total flux. The short shots are more frequent
($\sim$ 10--15 shots/s) and proportionally weaker, although their
contribution to the total flux is comparable to the longer
shots.

In general, the rapid time variability of the source 
X-ray flux is very similar to the low/hard state
of other Galactic black hole systems, such as Cyg~X-1, 
Nova Persei 1992, and GX~339-4.

The spectrum obtained by the PCA and HEXTE 
is clearly the hard/low state spectrum observed 
in many Galactic black hole binaries.  The overall power-law-with-
high-energy-cutoff shape can be approximated
by Comptonization models based on up-scattering of soft photons 
on energetic electrons in a hot plasma cloud.  In order
to fit the data, an additional component describing a spectrum 
reflected from cold material
with a strong iron fluorescent line must be included.
Both the equivalent width of the line and the intensity
of the reflected component are consistent with the assumption of
the reflection of hard X-ray emission from a cold, optically 
thick accretion disk.

To examine finer spectral changes we analyzed ratios of the raw spectra.  
This technique demonstrated an overall softening of the spectrum 
during the outburst, except for the last observation, 
which was obtained during the secondary maximum.  
At shorter time scales, we detect a
softening of the spectrum at higher flux levels.

An anti-correlation of fractional variability with energy 
is typical for Galactic black holes in their low spectral state
(e.g. \cite{nowak98a}, \cite{vihl95}, and this work), 
but a positive correlation is typical for neutron star systems 
in their low state (e.g. \cite{oliv98a}). 
Our analysis using RXTE archival data 
for several sources, confirmed this difference between 
black hole and neutron star binaries.  
This difference can be very useful for segregating neutron 
star binaries from black hole systems, which is otherwise
difficult with X-ray data only.

The research has made use of data obtained through the High Energy
Astrophysics Science Archive Research Center Online Service, provided 
by the NASA/Goddard Space Flight Center. 

MR is thankful to Dr.M.Gilfanov for helpful discussions.
KB is glad to acknowledge valuable comments by Prof.L.Titarchuk and
Dr.S.Brumby.  We are grateful to an anonymous referee for his/her
careful comments on the manuscript,
which helped us to improve the paper significantly.

\large
\clearpage


\begin{table}
\small
\caption{
RXTE observations of GS~1354-644.\label{obslog}}
\begin{tabular}{clcccccc}
\hline
\\
\#&Obs. ID& Date& Start time & TJD$^t$ &\multicolumn{2}{c}{Exposure, sec.}\\
& & & & &PCA&HEXTE$^d$\\
\hline
\\
1& 20431-01-01-01& 18/11/97& 00:28:32&10770.019& 1672& -\\
2& 20431-01-02-01,3& 19/11/97& 13:11:28& 10771.553& 5920  & 1812\\
3& 20431-01-03-00& 22/11/97& 09:39:28& 10774.402&6592& 2601\\
4& 20431-01-04-00& 05/12/97& 05:00:48& 10787.209&6809& 2053\\
5& 20431-01-05-00& 12/12/97& 19:21:20& 10794.806&6119& 1969\\
6& 30401-01-01-00& 27/12/97& 14:39:28& 10809.610&6516& 1896\\
7& 30401-01-02-00& 04/01/98& 17:47:12& 10817.741&7349& 2321\\
8& 30401-01-03-00& 12/01/98& 15:04:32& 10825.628&6355& 1932\\
9& 30401-01-04-00& 26/01/98& 12:07:28& 10839.505&6594& 2191\\
10& 30401-01-05-00& 17/11/98& 10:00:14& 11134.583&4109& 1597$^r$\\
\\
\hline
\end{tabular}
\begin{list}{}{}
\item[$^d$]- Dead time corrected exposure for each cluster of HEXTE detectors.
\item[$^r$]- This observation was in real-time format, so we could not
correct it for dead time 
\item[$^t$]- TJD=JD-244000.5, where JD is Julian Date or number of days since
Greenwich mean noon on Jan 1, 4713 B.C.
\end{list}
\end{table}

\begin{table}
\small
\caption{Best-fit parameters for power spectra approximation (3--60 keV, $10^{-3}$--20 Hz),\label{pds_par}}
\begin{tabular}{ccccccccc}
\hline
\\
Obs.&rms$_{total}$&$ Break_1$&$rms_1$&$Break_2$&$rms_2$&$Break_3$&$rms_3$&$\chi^2_{100dof}/100$\\
    & \%&$10^{-2}$ Hz&\%   &Hz   & \%&Hz&\% \\
\hline
\\
1&$ 39.2\pm  2.6$&$ 8.2\pm 0.9$&$ 27\pm 2$& $0.39\pm 0.23$&$ 16\pm6$&$ 2.31\pm 0.18$&$ 23\pm 1$&  1.4\\
2&$ 32.5\pm  1.1$&$ 2.5\pm 0.2$&$ 16\pm 1$& $0.36\pm 0.05$&$ 15\pm1$&$ 2.27\pm 0.10$&$ 20\pm 1$&  1.6\\

3&$ 31.8\pm  1.3$&$ 2.6\pm 0.2$&$ 14\pm 1$& $0.47\pm 0.06$&$ 16\pm1$&$ 2.55\pm 0.10$&$ 23\pm 1$&  1.2\\
4&$ 31.5\pm  1.2$&$ 5.8\pm 0.4$&$ 18\pm 1$& $0.72\pm 0.17$&$ 12\pm1$&$ 2.87\pm 0.16$&$ 23\pm 1$&  1.1\\
5&$ 29.8\pm  1.2$&$  7.5\pm 0.3$&$ 19\pm 1$& - & -&$ 2.30\pm 0.04$&$ 23\pm 1$&  1.5\\
6&$ 29.4\pm  1.7$&$  9.5\pm 0.3$&$ 19\pm 1$& - & -&$ 2.37\pm 0.05$&$ 23\pm 1$&  1.4\\
7&$ 28.9\pm  2.3$&$  9.6\pm 0.3$&$ 18\pm 2$& - & -&$ 2.48\pm 0.05$&$ 23\pm 1$&  1.2\\
8&$ 33.2\pm  2.6$&$ 10.1\pm 0.3$&$ 23\pm 2$& - & -&$ 2.50\pm 0.06$&$ 24\pm 2$&  1.3\\
9&$ 34.9\pm  4.1$&$  9.1\pm 0.3$&$ 27\pm 2$& - & -&$ 2.27\pm 0.07$&$ 22\pm 2$&  1.2\\
10&$<25$\\

\\
\hline
\end{tabular}
\begin{list}{}{}
\item[]- The power spectrum was approximated by a three component model. 
Each component is a function $\sim {1\over{1+({f\over{break_x}})^2}}$, 
corresponding to a simple exponential shot power spectrum. 
In the 6$^{th}$--9$^{th}$ observations the intermediate component was not 
significant.
\item[] - The systematic uncertainty of the background rate was included 
when calculating the errors in the $rms$ values of all components.
\end{list}
\end{table}

\begin{table}
\small
\caption{Spectral fit parameters for GS~1354--644.
\label{spectra_par}}

\begin{tabular}{cccccccccc}
\hline
\multicolumn{8}{c}{ $Pexrav$ (power law with cutoff + reflection) + Gauss$^d$}\\
\hline
Obs.\# & $\alpha$ & $E_{cutoff}$, keV & ${\Omega}/2{\pi}^a$&& EW, eV & F$^b$ & ${\chi}^2_{324 dof}/324$\\
\hline
\\
2--5 & $1.52\pm0.05$ & $123\pm8$ &$0.56\pm0.04$ &&  
$58\pm12$ & $59.5\pm0.5$ & 0.74\\
6--9 & $1.54\pm0.07$ & $230\pm20$ & $0.41\pm0.04$ && $53\pm12$ & $36.5\pm0.04$&0.99\\

\hline
\multicolumn{8}{c}{Broken power law with cutoff + Gauss$^d$}\\
\hline

Obs.\# & ${\alpha}_1$ & $E_{break}$, keV & ${\alpha}_2$ & $E_c$, keV & EW, eV & F$^b$ & ${\chi}^2_{323 dof}$/323\\
\hline
\\
2--5 & $1.41\pm0.02$ & $10.1\pm0.3$ & $1.12\pm0.02$ &$57\pm3$&$54\pm12$&$58.7\pm0.5$&0.75\\
6--9 & $1.46\pm0.02$ & $10.4\pm0.3$ & $1.18\pm0.02$ &$84\pm5$&$57\pm12$&$36.3\pm0.04$&0.92\\

\hline
\multicolumn{8}{c}{CompTT$^c$ + reflection + Gauss$^d$}\\
\hline

Obs.\#&$kT_e$, keV&$\tau$&$\Omega/2\pi^a$& & EW, eV & F$^b$ & $\chi^2_{324 dof}$/323\\
\hline
\\
2--5&$28\pm2$&$2.1\pm0.1$(disk)&$0.29\pm0.05$&&$63\pm15$&$57.5\pm0.5$&0.98\\
    &        &$4.8\pm0.2$(sphere)\\
6--9&$33\pm2$&$1.9\pm0.1$(disk)&$0.3\pm0.07$&&$57\pm12$&$35.5\pm0.5$&1.1\\
    &        &$4.3\pm0.2$(sphere)\\
\hline
\multicolumn{8}{c}{CompST(sphere) + reflection + Gauss$^d$}\\
\hline

Obs.\#&$kT_e$, keV&$\tau$&$\Omega/2\pi^a$&&EW, eV&F$^b$&$\chi^2_{324 dof}$/323\\
\hline
\\
2--5&$23.6\pm1.0$&$5.0\pm0.1$&$0.55\pm0.05$&&$65\pm12$&$54.3\pm0.8$&0.86\\
6--9&$28.8\pm1.0$&$4.5\pm0.1$&$0.33\pm0.05$&&$50\pm12$&$31.9\pm0.8$&0.92\\
\hline

\multicolumn{8}{c}{Power law}\\
\hline

Obs.\# & $\alpha$ & $F_{3-20 keV}$ & $F_{20-100 keV}$ &&&&$\chi^2_{43dof}$\\
\hline
10&$2.0\pm0.1$&$0.24\pm0.04$&$<1.2$&&&&0.64\\
\hline
\end{tabular}
\begin{list}{}{}
\item[$^a$]- The error bounds do not include the potentially important
effect of cross-calibration uncertainties between the PCA and HEXTE
\item[$^b$]- The source flux in the energy band 3--170 keV in units of 
$10^{-10}$ erg/s/cm$^2$. The HEXTE normalization was adjusted to the PCA
\item[$^c$]- For this model the optical depth was calculated for 
both sphere and disk geometries. The CompST model assumed a spherical 
geometry 
\item[$^d$]- The width of the Gaussian line was fixed at 0.1 keV;
$cos(\theta)$ ($\theta$ - inclination angle) for the reflection component 
was fixed at the value 0.45
\end{list}
\end{table}

\begin{table}
\small
\caption{Spectral fits to HEXTE data (20--170 keV).
\label{spectra_hexte_par}}

\begin{tabular}{cccc}
\hline
\multicolumn{4}{c}{ Power law with cutoff}\\
Obs.\# & $\alpha$ & $E_{cutoff}$, keV & ${\chi}^2_{282 dof}$\\
\hline
2--5&$1.13\pm0.04$&$66\pm4$&234\\ 
6--9&$1.05\pm0.05$&$80\pm8$&292\\
\hline
\end{tabular}
\begin{list}{}{}
\item[]- different HEXTE clusters (detectors) were fitted with 
different normalizations
\end{list}
\end{table}

\begin{table}
\small
\caption{Power-law fits to GS~1354--644 spectral ratios.
\label{spectratio_tbl}}
\begin{tabular}{ccccccc}
\hline
\\
\#Obs.&Flux/Flux$_{ref}$,\%&$\beta_1^*$,$\times10^{-2}$&$\chi^2_{44}$&Flux$_{max}$/Flux$_{min}$&$\beta_2^*$,$\times10^{-2}$&$\chi^2_{44}$\\
\hline
\\
1&$88\pm1$&$3.6\pm0.3$&0.74&$1.97\pm0.01$&$-5.7\pm0.8$&0.85\\
2&$88\pm1$&$4.6\pm0.3$&1.01&$2.35\pm0.01$&$-5.7\pm0.8$&1.00\\
3&$98\pm1$&$2.7\pm0.2$&0.70&$1.86\pm0.01$&$-5.7\pm0.4$&1.13\\
4& -      &      -     & -  &$1.83\pm0.01$&$-6.8\pm0.7$&1.18\\
5&$89\pm1$&$-0.3\pm0.2$&0.73&$1.95\pm0.01$&$-6.2\pm1.0$&0.91\\
6&$69\pm1$&$-1.7\pm0.3$&0.72&$1.59\pm0.01$&$-3.6\pm0.7$&0.95\\
7&$54\pm1$&$-1.7\pm0.2$&1.01&$1.57\pm0.01$&$-2.9\pm0.7$&1.16\\
8&$51\pm1$&$-1.9\pm0.2$&0.75&$1.41\pm0.01$&$-2.3\pm0.7$&0.79\\
9&$59\pm1$&$8.4\pm0.5$&1.06&$1.70\pm0.01$&$-22\pm2$&1.02\\
10&$2.2\pm0.3$&$-59\pm6$&0.55&-&-&-\\
\\
\hline
\end{tabular}
\begin{list}{}{}
\item[$^*$]-- $\beta_1$ is the power-law index of the spectral ratio 
for an individual observation normalized to observation \#4. 
$\beta_2$ is the same parameter for ratio of high-flux and
low-flux spectra within a single observation.
The power-law indices $\beta_1$ and $\beta_2$ can be measured 
with very high accuracy as long as the instrument response is stable. 
Observations of the Crab nebula show that the slope of 
the spectrum is constant to  $\sim6\times10^{-3}$
for up to 5 months.
\end{list}
\end{table}

\begin{table}
\small
\caption{Energy dependence of the rms fractional variability
for Galactic X-ray binaries observed in low/hard state,
expressed as a power law slope in the $\sim$3--15 keV energy range.
\label{rms_en}}
\begin{tabular}{cccc}
\hline
\\
Source & Type & Date & PL slope\\
\hline
\\
GS~1354--644 & BH & Nov 19, 1997 & $-0.07\pm0.01$\\ 
Cyg~X-1 & BH & Jun 26, 1997 & $-0.05\pm0.005$\\ 
GX~339-4 & BH & Sep 19, 1997&$-0.035\pm0.01^a$\\
\hline
Terzan~2 & NS & Nov 1996 & $+0.25\pm0.04^b$\\ 
GX~354-0 & NS & Mar 3, 1996& $+0.08^c$\\ 
SAX~J1808.4-3658& NS & Apr 13, 1998 & $+0.16\pm0.03$\\
4U1608-522& NS & Dec 27, 1996& $+0.6\pm0.3$\\
4U0614+091& NS & Jan 25, 1997 & $+0.20\pm0.04$\\
4U1705-44& NS  & Mar 29, 1997 & $+0.13\pm0.05$\\
SLX~1735-269& NS & Feb-May, 1997& $+0.10\pm0.17$\\
\\
\hline
\end{tabular}
\begin{list}{}{}
\item[$^a$]- \cite{nowak98c}
\item[$^b$]- \cite{oliv98a}
\item[$^c$]- No confidence interval is quoted, 
because the rms variability dependence on energy 
does not fit a simple power-law (see Fig.~4a)
\end{list}
\end{table}

\end{document}